\documentclass[fleqn]{svmult}
\usepackage{epsfig}

\def\lsim{\mathrel{\raise.2ex\hbox{$<$}\hskip-.8em\lower.9ex\hbox{$\sim$}}}
\def\gsim{\mathrel{\raise.2ex\hbox{$>$}\hskip-.8em\lower.9ex\hbox{$\sim$}}}

\let\agt=\gsim

\def\Re{\;{\cal R}\!e}
\def\Im{\;{\cal I}\!m}

\begin{document}
\font\fortssbx=cmssbx10 scaled \magstep1

\title*{\hbox to \hsize{%
%\special{psfile=uwlogo.ps hscale=6000 vscale=6000 hoffset=-12 voffset=-2}
%\hskip.35in \raise.1in
\hbox{\fortssbx University of Wisconsin - Madison}
\hfill\vtop{\normalsize\hbox{\bf MADPH-00-1180}
                \hbox{\rm May 2000}
                \hbox{\hfil}\hbox{\hfil}} }
Neutrino Mixing Schemes}

\author{V. Barger$^a$ and K. Whisnant$^b$}

\institute{$^a$Department of Physics, University of Wisconsin, Madison, WI 53706\hfill\break
$^b$Department of Physics and Astronomy, Iowa State University,
Ames, IA 50011}
\maketitle

\renewcommand{\thefootnote}{\relax}

\footnotetext{To appear in ``Current Aspects of Neutrino Physic", ed. by D.~Caldwell (Springer-Verlag, Hamburg, 2000)}

\section{Introduction}

A revolution in our understanding of the neutrino sector is underway, driven by
observations that are interpreted in terms of changes in neutrino flavors as
they propagate. Since neutrino oscillations occur only if neutrinos are
massive, these phenomena indicate physics beyond the Standard Model. With the
present evidence for oscillations from atmospheric, solar and accelerator data,
we are already able to begin to make strong inferences about the mass spectrum
and mixings of neutrinos. Theoretical efforts to achieve a synthesis have
produced a variety of models with differing testable consequences. A
combination of particle physics, nuclear physics and astrophysics is needed for
a full determination of the fundamental properties of neutrinos. This article
reviews what has been achieved thus far and the future prospects for
understanding the nature of neutrino masses and mixing.

\section{Two-Neutrino Analyses}

In a model with two neutrinos, the probability for a given neutrino flavor
$\nu_\alpha$ oscillating into $\nu_\beta$ in a vacuum is
\begin{equation}
P(\nu_\alpha\to\nu_\beta) = \sin^22\theta \sin^2 1.27 {\delta m^2 L\over E} \,,
\end{equation}
where $\theta$ is the two-neutrino mixing angle, $\delta m^2$ is the
mass-squared difference of the two mass eigenstates in eV$^2$, $L$ is the
distance from the neutrino source to the detector in kilometers, and $E$ is the
neutrino energy in GeV.

\subsection{Atmospheric neutrinos}

The atmospheric neutrino experiments determine the ratios
\begin{eqnarray}
{N_\mu\over N_\mu^0} &=& \alpha \left[ \left< P(\nu_\mu\to\nu_\mu) \right> + r
\left< P(\nu_e\to\nu_\mu \right> \right] \,, \\
{N_e\over N_e^0} &=& \alpha \left[ \left< P(\nu_e\to\nu_e \right> + r^{-1}
\left< P(\nu_\mu\to\nu_e \right> \right] \,,
\end{eqnarray}
where $N_e^0$ and $N_\mu^0$ are the expected numbers of atmospheric $e$ and
$\mu$ events, respectively, in the absence of oscillations, $r\equiv{N_e^0\over N_\mu^0}$, $\left<~\right>$
indicates an average over the neutrino spectrum, and $\alpha$ is an overall
neutrino flux normalization. Atmospheric neutrino data have generally
indicated that the expected number of muons detected is suppressed relative to
the expected number of electrons~\cite{atmos}; this suppression can be
explained via neutrino oscillations~\cite{oldatmos}.

The atmospheric data also indicate that $N_e/N_e^0$ is relatively flat
with zenith angle, while $N_\mu/N_\mu^0$ decreases with increasing
zenith angle (i.e., with longer oscillation distance). Assuming
$\nu_\mu\to\nu_\tau$ oscillations, the favored two-neutrino parameters
are~\cite{SuperKatmos}
\begin{eqnarray}
\delta m^2 &=& 3.5\times 10^{-3}\rm\,eV^2\quad
(1.5\mbox{--}7\times10^{-3}\,eV^2) \,, \label{eq:dm2}\\
\sin^22\theta &=& 1.00 \quad (0.80\mbox{--}1.00) \,;
\end{eqnarray}
the 90\% C.L. allowed ranges are given in parentheses. The absolute
normalization of the electron data indicates $\alpha\simeq1.18$, which is
within the theoretical uncertainties~\cite{atmosth}. The flatness versus
$L$ of the electron data implies
that simple $\nu_\mu\to\nu_e$ oscillations are strongly disfavored. Large
amplitude ($\sin^22\theta>0.2$) $\nu_\mu\to\nu_e$ oscillations are also
excluded by the CHOOZ reactor data~\cite{CHOOZ} for $\delta m^2_{\rm
atm} \agt 10^{-3}\rm\,eV^2$.

It is also possible that atmospheric $\nu_\mu$ are oscillating into
sterile neutrinos. However, measurements of the upgoing zenith angle
distribution and $\pi^0$ production disfavor this possibility~\cite{nusatm,jglchap}.

\subsection{Solar neutrinos}

For the $^{37}$Cl~\cite{Cl} and $^{71}$Ga~\cite{Ga} experiments, the expected
number of neutrino events is
\begin{equation}
N = \int \sigma P(\nu_e\to\nu_e) (\beta \phi_{\rm B} + \phi_{\rm non{-}B} )
dE\,,
\end{equation}
where we allow an arbitrary normalization factor $\beta$ for the
$^8$B neutrino flux since its normalization is not certain. For
the Kamiokande~\cite{Kam} and Super-Kamiokande~\cite{SuperKsolar}
experiments the interaction is $\nu e \to \nu e$ and the outgoing
electron energy is measured. The number of events per unit of electron
energy is then
\begin{eqnarray}
{dN\over dE_e} &=& \beta \int \left\{ {d\sigma_{CC}\over dE'_e} P(\nu_e\to\nu_e)
+ {d\sigma_{NC}\over dE'_e} \left[ 1 - P(\nu_e\to\nu_e) \right] \right\}
\nonumber\\
&& \hspace{5em} {}\times G\left(E'_e, E_e\right) \phi_{\rm B} dE_\nu dE'_e \,,
\end{eqnarray}
where $d\sigma_{CC}/dE'_e\ (d\sigma_{NC}/dE'_e)$ is the charged-current
(neutral-current) differential cross section for an incident neutrino energy
$E_\nu$ and $G(E'_e, E_e)$ is the probability that an electron of energy $E'_e$
is measured as having energy $E_e$. The neutrino fluxes are taken from the
standard solar model (SSM)~\cite{SSM}. If $\nu_e$ oscillates into a
sterile neutrino, $\sigma_{NC}=0$.

For two-neutrino vacuum oscillations (VO) of $\nu_e$ into either $\nu_\mu$
or $\nu_\tau$~\cite{vacuum}, the solar data indicate~\cite{bw98vac,bks98}
\begin{eqnarray}
\delta m^2 &=& 7.5\times10^{-11}\rm\,eV^2 \,,\label{eq:dm2-b}\\
\sin^2 2\theta &=& 0.91 \,,
\end{eqnarray}
although there are also regions near $\delta m^2 = 2.5\times10^{-10},\
4.4\times10^{-10},\ \rm and\ 6.4\times10^{-10}\,eV^2$ that also give acceptable
fits.

For two-neutrino MSW oscillations~\cite{MSW} of $\nu_e$ into either
$\nu_\mu$ or $\nu_\tau$, there are three possible solution
regions~\cite{bks98}. The small-angle MSW (SAM) solution is
\begin{eqnarray}
\delta m^2 &=& 7.5\times10^{-6}\rm\,eV^2 \,,\label{eq:dm2-c}\\
\sin^22\theta &=& 0.01 \,,\\
\noalign{\hbox{and the large-angle MSW (LAM) solution is}}
\delta m^2 &\sim& 10^{-5}\rm\,eV^2 \,,\label{eq:dm2-d}\\
\sin^22\theta &=& 0.6\mbox{--}0.9 \,.\\
\noalign{\hbox{There is also a low $\delta m^2$ MSW (LOW) solution}}
\delta m^2 &\sim& 10^{-7}\rm\,eV^2 \,,\label{eq:dm2-e}\\
\sin^22\theta &=& 0.6\mbox{--}0.9
\end{eqnarray}
although it is less favored. Note that all solutions except for
small-angle MSW have large mixing in the solar sector.

Two-neutrino oscillations of $\nu_e$ into a sterile neutrino are somewhat
disfavored because sterile neutrinos do not have a NC interaction, which tends
to make oscillation predictions for $^{37}$Cl and SuperK similar, contrary to
experimental evidence. Future measurements of NC interactions in the SNO
detector~\cite{SNO} will provide a more thorough test for oscillations of solar
$\nu_e$ into sterile neutrinos.

SuperK and SNO will also provide an improved measurement of the $^8$B
neutrino spectrum in the near future, which should help distinguish the
various solar scenarios. The Borexino~\cite{borexino} and ICARUS~\cite{icarus} experiments will
provide a measurement of the $^7$Be neutrinos, and could detect the
strong seasonal dependence that exists for $^7$Be neutrinos in many VO
models~\cite{seasonal}.

\subsection{LSND}

There are also indications of neutrino oscillations from the LSND accelerator
experiment~\cite{LSND}. Their data suggest $\nu_\mu\to\nu_e$ oscillations with
two-neutrino parameters,
\begin{equation}
0.3{\rm\ eV^2} \le \delta m^2 = {0.03{\rm\ eV^2}\over (\sin^22\theta)^{0.7}}
\le 2.0\rm\ eV^2 \,.
\end{equation}
Values of $\delta m^2$ above 10~eV$^2$ are also allowed for
$\sin^22\theta\simeq0.0025$\cite{caldwell}. The future MiniBooNE experiment~\cite{miniboone} is
expected to either confirm or refute the LSND result. 

\section{Global Analyses}

\looseness=-1
A complete description of all neutrino data requires a model that can
explain all of the oscillation phenomena simultaneously. Since each of
the three types of oscillation evidence (atmospheric, solar, LSND)
requires a distinct $\delta m^2$ scale, and since $N$ neutrinos have
only $N-1$ independent mass-squared differences, four neutrinos are
needed to completely explain all of the data.  If one of the types of
neutrino data is ignored, then it is in principle possible to explain
the remaining data with a three-neutrino model. Because the LSND results
have yet to be confirmed by another experiment, three-neutrino models
are generally used in the context of describing the atmospheric and
solar data.

\subsection{Three neutrinos}

\begin{figure}[h]
\centering\leavevmode
\epsfxsize=4in\epsffile{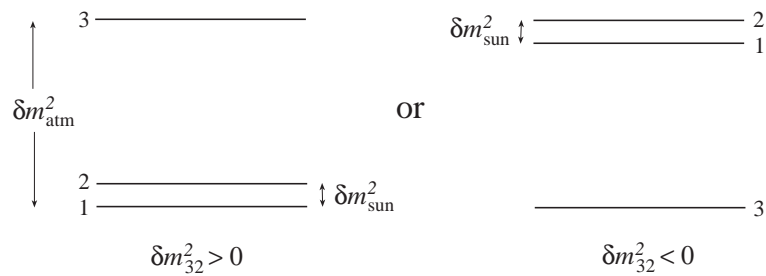}

\caption[]{Two possible patterns of neutrino masses that can explain the atmospheric and solar anomalies.\label{fig:3nu-mass}}
\end{figure}

If the atmospheric and solar data are to be described by a
three-neutrino model, there are two distinct mass-squared difference
scales [compare Eq.~(\ref{eq:dm2}) with Eqs.~(\ref{eq:dm2-b}),
(\ref{eq:dm2-c}), (\ref{eq:dm2-d}) and (\ref{eq:dm2-e})]. The two possible patterns of three-neutrino masses are illustrated in Fig.~\ref{fig:3nu-mass}.
We assume
without loss of generality that $|m_1|<|m_2|<|m_3|$, and that $\delta
m_{21}^2 \equiv \delta m_{\rm sun}^2$ and $\delta m_{31}^2 \simeq \delta
m_{32}^2 \equiv \delta m_{\rm atm}^2 \gg \delta m_{\rm sun}^2$. Then the
off-diagonal vacuum oscillation probabilities in a three-neutrino model
are~\cite{bww2}
\begin{eqnarray}
P(\nu_e\to\nu_\mu) &=& 4\left| U_{e3} U_{\mu3}^*\right|^2 \sin^2\Delta_{\rm
atm} 
 -4\Re \left\{ U_{e1} U_{e2}^* U_{\mu1}^* U_{\mu2} \right\} \sin^2\Delta_{\rm
sun} \nonumber\\
&& -2 J\sin 2 \Delta_{\rm sun} \,, \\
P(\nu_e\to\nu_\tau) &=& 4\left| U_{e3} U_{\tau2}^* \right|^2 \sin^2\Delta_{\rm
atm} 
 - 4\Re \left\{ U_{e1} U_{e2}^* U_{\tau1}^* U_{\tau2} \right\}
\sin^2\Delta_{\rm sun} \nonumber\\
&& + 2J\sin 2\Delta_{\rm sun} \,, \\
P(\nu_\mu\to\nu_\tau) &=& 4\left| U_{\mu3} U_{\tau3}^* \right|^2
\sin^2\Delta_{\rm atm} 
 -4\Re \left\{ U_{\mu1} U_{\mu2}^* U_{\tau1}^* U_{\tau2} \right\}
\sin^2\Delta_{\rm sun} \nonumber\\
&&-2J\sin 2\Delta_{\rm sun} \,, 
\end{eqnarray}
where $U$ is the neutrino mixing matrix (in the basis where the charged-lepton
mass matrix is diagonal), $\Delta_j \equiv 1.27(\delta m_j^2 /{\rm eV^2})
(L/{\rm km}) (E/\rm GeV)$ and $J = \Im\left\{ U_{e2} U_{e3}^* U_{\mu2}^*
U_{\mu3}\right\}$ is the CP-violating invariant~\cite{jarlskog}.
For a discussion of CP violating effects in neutrino oscillations see Refs.~\cite{4nuCP,deruj,cpv}

%% orig place of fig.1 call

The matrix elements $U_{\alpha j}$ are the mixing between flavor ($\alpha =
e,\mu,\tau$) and the mass ($j=1,2,3$) eigenstates. The CP-odd term changes sign
under reversal of the oscillating flavors, or if neutrinos are replaced by
anti-neutrinos. For either Dirac or Majorana neutrinos we choose the following
parametrization for $U$ to describe neutrino oscillations
\begin{equation}
U = \left(\begin{array}{ccc}
c_{13} c_{12}& c_{13} s_{12}& s_{13} e^{-i\delta}\\
-c_{23} s_{12} -s_{13} s_{23} c_{12}e^{i\delta}& c_{23} c_{12}-s_{13}s_{23}
s_{12}e^{i\delta}& c_{13} s_{23}\\
s_{23}s_{12}-s_{13} c_{23} c_{12}e^{i\delta}& -s_{23}c_{12} -s_{13}c_{23}
s_{12}e^{i\delta}& c_{13} c_{23}
\end{array}\right) \,,
\end{equation}
where $c_{jk}\equiv \cos\theta_{jk}$ and $s_{jk}\equiv \sin\theta_{jk}$. For
Majorana neutrinos, $U$ contains two further phase factors, but these do not enter into oscillation
phenomena.

For the oscillation of neutrinos in atmospheric and long-baseline experiments
with $L/E \agt 10^2\,$km/GeV, the $\Delta_{\rm sun}$ terms are negligible and
the relevant vacuum oscillation probabilities are
\begin{eqnarray}
P(\nu_e\to\nu_\mu) &=& s_{23}^2 \sin^22\theta_{13} \sin^2 \Delta_{\rm atm}\,,
\\
P(\nu_e\to\nu_\tau)&=& c_{23}^2 \sin^22\theta_{13} \sin^2\Delta_{\rm atm}\,,
\\
P(\nu_\mu\to\nu_\tau) &=& c_{13}^4 \sin^22\theta_{23} \sin^2\Delta_{\rm atm}\,.
\label{eq:pmutau}
\end{eqnarray}
The diagonal oscillation probabilities for a given neutrino species can be
found by subtracting all the off-diagonal probabilities involving that
species from unity.

For neutrinos from the sun, $L/E \sim 10^{10}\,$km/GeV, and the
$\sin^2\Delta_{\rm atm}$ terms oscillate very rapidly, averaging to 1/2. Then
the observable oscillation probability in a vacuum is
\begin{equation}
P(\nu_e\to\nu_e) = 1 - {1\over2} \sin^2 2\theta_{13} - c_{13}^4
\sin^22\theta_{12} \sin^2 \Delta_{\rm sun} \,.
\label{eq:P3solar}
\end{equation}
When $\theta_{13} = 0$ (i.e., $U_{13}=0$), the oscillations of
atmospheric and long-baseline neutrinos decouple from those of solar
neutrinos: at the $\delta m_{\rm atm}^2$ scale, there are pure
$\nu_\mu\to\nu_\tau$ oscillations with amplitude $\sin^22\theta_{23}$,
with no admixture of $\nu_e$, and at the $\delta m_{\rm sun}^2$ scale
the $\nu_e$ oscillations are described by a simple two-neutrino formula
with amplitude $\sin^22\theta_{12}$. Then the two-neutrino parameters
for atmospheric and solar oscillations can be adopted directly in the
three-neutrino case.

If $\theta_{13}\neq0$, then $\nu_e$ will participate in atmospheric and
long-baseline oscillations. As discussed earlier, pure $\nu_\mu\to\nu_e$
oscillations at the $\delta m_{\rm atm}^2$ scale are strongly disfavored
by the atmospheric data, but some small amount of $\nu_\mu\to\nu_e$ is
still allowed.  The CHOOZ reactor experiment measures $\bar\nu_e$
survival, and is sensitive to oscillations between $\bar\nu_e$ and
$\bar\nu_\mu$ for $\delta m_{\rm atm}^2 > 10^{-3}\rm\,eV^2$. The
combined data from atmospheric experiments and CHOOZ favor
$\sin\theta_{13}=0$ (i.e., pure $\nu_\mu\to\nu_\tau$ oscillations in
the atmosphere) and suggest that $\sin\theta_{13}<0.3$~\cite{bww2}.

The solar data also allow solar neutrinos to mix maximally, or nearly
maximally. If we require both atmospheric and solar oscillations to be
maximal, there is a unique three-neutrino solution to the neutrino
mixing matrix~\cite{bimaximal}. This ``bimaximal" mixing corresponds to
$\theta_{13}=0$ and $|\theta_{12}|=|\theta_{23}|=\pi/4$, and is a special
case of the decoupled solution for atmospheric and solar neutrinos. One
interesting aspect of the bimaximal solution is that the solar $\nu_e$
oscillations are 50\% into $\nu_\mu$ and 50\% into $\nu_\tau$, although
the flavor content of the $\nu_e$ oscillation is not observable in solar
experiments. Further properties of the bimaximal and nearly bimaximal
solutions and models that give rise to such solutions are discussed in Ref.~\cite{bimaximal}.

\subsection{Four neutrinos}

\begin{figure}
\centering\leavevmode
\epsfxsize=2in\epsffile{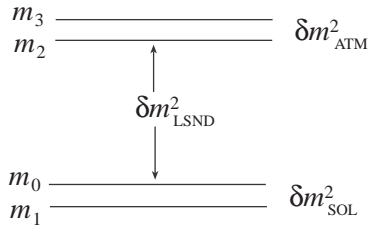}

\caption[]{Required mass pattern of two separated pairs to account for the LSND, atmospheric, and solar data; the locations of $\delta m_{\rm ATM}^2$ and $\delta m_{\rm SUN}^2$ may be interchanged.\label{fig:4nu-mass}}
\end{figure}

As discussed earlier, four neutrinos are necessary to completely
describe the atmospheric, solar and LSND results. A fourth light
neutrino must be sterile since it is not observed in $Z$
decays~\cite{zwidth}. There must be three separate mass--squared
scales which must satisfy the hierarchy $\delta m^2_{sun} \ll \delta
m^2_{atm} \ll \delta m^2_{LSND}$. In the three--neutrino case the
relation $\delta m^2_{sun} \ll \delta m^2_{atm}$ leads to only one
fundamental type of mass structure, in which one heavier mass is
separated from two lighter, nearly degenerate states (or vice versa).
In the four--neutrino case, however, there are two distinct types of
mass structures: one heavier mass separated from three lighter, nearly
degenerate states, or vice versa (which we will refer to as the $1+3$
spectrum), or two nearly degenerate mass pairs separated from each other
(the $2+2$ spectrum). In each case, the largest separation scale is
determined by the LSND scale. It can be shown~\cite{BGG,4nuvar} that
only the $2+2$ spectrum is completely consistent with the positive
oscillation signals in the solar, atmospheric and LSND experiments, and
the negative results from the BUGEY reactor~\cite{Bugey} and CDHS
accelerator~\cite{CDHS} experiments. Therefore our discussions below
assume the $2+2$ case, which is illustrated in Fig.~\ref{fig:4nu-mass}. Here we will assume that the mass splitting of the
heavier pair drives the atmospheric oscillations, the mass splitting of
the lighter pair drives the solar oscillations, and the separation of
the mass pairs drives the LSND oscillations.  Sterile neutrinos are also of interest in explaining $r$-process nucleosynthesis via supernova explosions (see e.g. Ref.~\cite{caldwell}).

The vacuum neutrino flavor oscillation probabilities, for an initially
produced $\nu_\alpha$ to a finally detected $\nu_\beta$, can be
written~\cite{4nuCP}
\begin{equation}
P(\nu_\alpha \rightarrow \nu_\beta) =
\delta_{\alpha\beta} - \sum_{j<k} \left[
4 {\rm~Re}(W^{jk}_{\alpha\beta}) \sin^2\Delta_{kj}
- 2 {\rm~Im}(W^{jk}_{\alpha\beta}) \sin2\Delta_{kj} \right] \,,
\label{eq:genprob}
\end{equation}
where
\begin{equation}
W^{jk}_{\alpha\beta} =
U_{\alpha j} U_{\alpha k}^* U_{\beta j}^* U_{\beta k} \,,
\label{eq:W}
\end{equation}
and $\Delta_{kj} \equiv 1.27 \delta m^2_{kj} L/E$. We assume that
there are four mass eigenstates $m_0,m_1, m_2,m_3,m_4$, which are most
closely associated with the flavor states $\nu_s$, $\nu_e$, $\nu_\mu$,
and $\nu_\tau$, respectively. The solar oscillations are driven by
$\delta m^2_{01}$, the atmospheric oscillations by $\delta m^2_{32}$,
and the LSND oscillations by $\delta m^2_{02} \simeq \delta m^2_{03}
\simeq \delta m^2_{12} \simeq \delta m^2_{13}$. Hence, the solar
oscillations are primarily $\nu_e\to\nu_s$ and the atmospheric
oscillations are primarily $\nu_\mu\to\nu_\tau$. For oscillations of
solar $\nu_e$ to sterile neutrinos, the solar data disfavors large
mixing; hence the most likely solar solution is MSW small mixing. It is
also possible to
reverse the roles of $\nu_s$ and $\nu_\tau$; however, current data disfavor
oscillations of atmospheric $\nu_\mu$ to sterile neutrinos~\cite{nusatm}.

The $4\times4$ mixing matrix $U$ may be parametrized by 6 mixing angles
($\theta_{01}$, $\theta_{02}$, $\theta_{03}$, $\theta_{12}$,
$\theta_{13}$, $\theta_{23}$) and 6 phases ($\delta_{01}$,
$\delta_{02}$, $\delta_{03}$, $\delta_{12}$, $\delta_{13}$,
$\delta_{23}$); only three of the phases are observable in neutrino
oscillations.  A complete analysis of the four-neutrino mixing matrix
is quite complicated. However, the smallness of the mixing indicated by
the LSND results suggests that $\nu_e$ does not mix strongly with the two
heavier states, i.e., that $\theta_{12}$ and $\theta_{13}$ are small.
Similarly, one can assume that the other light state does not mix
strongly with the heavier states, i.e., $\theta_{02}$ and $\theta_{03}$
are also small. This situation occurs
naturally in the explicit four-neutrino models in the literature. Then
after dropping terms second order in small mixing angles, $U$ takes the
form~\cite{4nuCP}
\begin{equation}
U = \left( \begin{array}{cccc}
c_{01} & s_{01}^* & s_{02}^* & s_{03}^*
\\
&&&\\
-s_{01} & c_{01} & s_{12}^* & s_{13}^*
\\
&&&\\
-c_{01}(s_{23}^*s_{03}+c_{23}s_{02})
& -s_{01}^*(s_{23}^*s_{03}+c_{23}s_{02})
& c_{23}
& s_{23}^*
\\
+s_{01}(s_{23}^*s_{13}+c_{23}s_{12})
& -c_{01}(s_{23}^*s_{13}+c_{23}s_{12})
&&\\
&&&\\
c_{01}(s_{23}s_{02}-c_{23}s_{03})
& s_{01}^*(s_{23}s_{02}c_{23}s_{03})
& -s_{23}
& c_{23}
\\
-s_{01}(s_{23}s_{12}-c_{23}s_{13})
& +c_{01}(s_{23}s_{12}-c_{23}s_{13})
&&\\
&&&\\
\end{array} \right) \,,
\label{eq:genU2}
\end{equation}
where $c_{jk} \equiv \cos\theta_{jk}$ and $s_{jk} \equiv
\sin\theta_{jk}e^{i\delta{jk}}$. We see that under these conditions, $U$
has approximately block diagonal form. The mixing of solar neutrinos is
due to $\theta_{01}$ and the mixing of atmospheric neutrinos is due to
$\theta_{23}$; the values for these mixing angles are essentially given
by the two-neutrino fits in Sec.~2. Both vacuum and MSW solar
oscillation solutions are allowed; for MSW oscillations to occur in the
sun requires $m_0>m_1$. The mixing that drives the LSND
oscillations is due to $\theta_{12}$ and $\theta_{13}$; the effective
amplitude of the $\nu_\mu\to\nu_e$ oscillations in LSND is
\begin{equation}
\sin^22\theta_{LSND} \simeq 4|s_{12}c_{23} + s_{13}s_{23}^*|^2 \,.
\label{eq:LSNDamp}
\end{equation}

The assumption that $\theta_{02}$ and $\theta_{03}$ are small is not
required by current data. However, most explicit models have the
approximate form given by Eq.~(\ref{eq:genU2}); see Sec.~4.2. If in fact
$\theta_{02}$ and $\theta_{03}$ are not small, $\nu_e$ oscillations to
the flavor associated with the other light state ($\nu_s$ or $\nu_\tau$)
are possible at the LSND $L/E$ scale with an amplitude of the same order
as the LSND oscillation amplitude.

\section{Consequences for Masses and Mixings}

\subsection{Three-neutrino models}

The atmospheric and solar data put restrictions on the neutrino mixing
angles and mass-squared differences, but do not at all constrain the
absolute neutrino mass scale, which must be determined by other
methods. The freedom to choose the mass scale allows a wide variety of
possible mass matrix structures, even for the same mass-squared
differences and mixing.

The matrix $U$ that relates neutrino flavor eigenstates to neutrino mass
eigenstates depends in general on mixing in both the neutrino and
charged lepton sector. If $U_\ell$ diagonalizes the charged lepton mass
matrix and $U_\nu$ the neutrino mass matrix, then $U = U_\ell^\dagger
U_\nu$. Except where stated otherwise, in our discussions here we will
work in the basis where the charged lepton mass matrix is diagonal, so
that $U = U_\nu$. In general, the neutrino mass matrix in the flavor
basis can then be written $M_{\alpha\beta} = \sum_j U_{\alpha j} m_j
U_{\beta j}$ for Majorana neutrinos or $M_{\alpha\beta} = \sum_j
U_{\alpha j} m_j U_{\beta j}^*$ for Dirac neutrinos (these are the same
if $CP$ is conserved, i.e., when $U$ is real).

As an example of the different mass matrix structures that are possible,
we consider the case when at least one of the masses is much smaller
than the largest mass. Then there is one type of mass matrix of the form
$M = M_0 + O(\delta m^2_{jk})$ (up to trivial sign changes) that can
lead to maximal mixing of atmospheric neutrinos:
\begin{equation}
M_0 = {m\over2} \left(
\begin{array}{ccc} 0 & 0 & 0 \\ 0 & a & b \\ 0 & b & a \end{array}
\right) \,,
\label{eq:M1}
\end{equation}
where $a,b \sim 1$. If $a=b$, then there is only one large mass ($m_1,
m_2 \ll m_3 \equiv m$) and the form of Eq.~(\ref{eq:M1}) automatically
fixes $\theta_{23} \simeq \pi/4$ and $\theta_{13} \simeq 0$; the
$O(\delta m^2_{jk})$ terms determine $\theta_{12}$. Bimaximal mixing
($\theta_{12} \simeq \pi/4$) is obtained if the leading
$O(\delta m^2_{jk})$ terms have the form
\begin{equation}
\Delta M = \epsilon \left(
\begin{array}{ccc} 0 & -1 & 1 \\ -1 & 0 & 0 \\ 1 & 0 & 0 \end{array}
\right) \,,
\label{eq:M4}
\end{equation}
where $\epsilon \ll m$; subleading $O(\delta m^2_{jk})$ terms are then
needed to split $m_1$ and $m_2$. If $a \ne b$ in Eq.~(\ref{eq:M1}), then
there are two large masses with $\theta_{23} \simeq \pi/4$ and
$\theta_{13} \simeq 0$. Small $\theta_{12}$, appropriate for
the small-angle MSW solar solution, is achieved if the leading
$O(\delta m^2_{jk})$ terms then have the form
\begin{equation}
\Delta M = \epsilon \left(
\begin{array}{ccc} 0 & 1 & 1 \\ 1 & 0 & 0 \\ 1 & 0 & 0 \end{array}
\right) \,;
\label{eq:M5}
\end{equation}
see Ref.~\cite{ramond} for an example of a GUT model that has this
form.

In the situation where all masses are approximately degenerate,
$m \equiv |m_1| \simeq |m_2| \simeq |m_3| \gg
\delta m^2_{jk}$, there are three different types of mass matrices of
the form $M = M_0 + O(\delta m^2_{jk})$ (up to trivial changes in sign)
that can lead to bimaximal mixing, depending on the relative signs of
the $m_j$:
\begin{equation}
M_0 =
m \left(
\begin{array}{ccc} 0 & -{1\over\sqrt2} & {1\over\sqrt2} \\
-{1\over\sqrt2} & {1\over2} & {1\over2} \\
{1\over\sqrt2} & {1\over2} & {1\over2} \end{array}
\right) \,, \ {\rm~or~} \ 
m \left(
\begin{array}{ccc} 1 & 0 & 0 \\ 0 & 0 & 1 \\ 0 & 1 & 0 \end{array}
\right) \,, \ {\rm~or~} \ 
m \left(
\begin{array}{ccc} 1 & 0 & 0 \\ 0 & 1 & 0 \\ 0 & 0 & 1 \end{array}
\right) \,.
\label{eq:M2}
\end{equation}
The requirement from neutrinoless double beta decay that the
$\nu_e\nu_e$ element of the neutrino mass matrix be small (described
below) implies that only
the first case is allowed for Majorana neutrinos. The form of $M_0$
gives three degenerate neutrinos of mass $m$ and fixes two combinations
of mixing angles ($c_{13} s_{12} \simeq -1/\sqrt2$ and $c_{23}c_{12} -
s_{13}s_{23}s_{12} \simeq 1/2$), while the remaining degree of freedom
among the mixing angles and the neutrino mass splittings are determined
by the $O(\delta m^2_{jk})$ terms. If the leading $O(\delta m^2_{jk})$
terms are proportional to the mass matrix in Eq.~(\ref{eq:M1}) with
$a=b$, $m_3$ is split from $m_1$ and $m_2$, $\theta_{13} \simeq 0$ and
bimaximal mixing is obtained. Subleading $O(\delta m^2_{jk})$ terms then
provide the splitting between $m_1$ and $m_2$.

A different mixing scheme occurs if the neutrino mass matrix is
approximately proportional to unity and the charged lepton mass
matrix is close to the so-called democratic form~\cite{demo}
\begin{equation}
M_\ell =
{m_\ell\over 3} \left(
\begin{array}{ccc} 1 & 1 & 1 \\ 1 & 1 & 1 \\ 1 & 1 & 1 \end{array}
\right) \,;
\label{eq:M3}
\end{equation}
then there is one large eigenvalue $m_\ell \simeq m_\tau$ and two
constraints on the flavor mixing angles, $c_{13}c_{23} \simeq
1/\sqrt3$ and $s_{23}s_{12} - s_{13}c_{23}c_{12} \simeq 1/\sqrt3$. If
a small perturbation is added to the lower right element of $M_\ell$,
the muon gets mass and $\theta_{13}$ is constrained to be approximately
zero; then there is maximal mixing in the solar sector
($\sin^22\theta_{12} = 1$) and nearly maximal mixing in the atmospheric
sector ($\sin^22\theta_{23} = 8/9$)~\cite{demo}. Additional
perturbations to the diagonal elements of the charged lepton and
neutrino mass matrices are then needed to give the electron-muon and
neutrino mass splittings, respectively.

An $SO(10)$ SUSY GUT model with a minimal Higgs sector can provide
large $\nu_\mu$--$\nu_\tau$ mixing for atmospheric neutrinos,
and can accommodate either small or large mixing of solar
neutrinos~\cite{albright}.  
Other possible neutrino mass matrix textures are discussed in
Refs.~\cite{textures}, \cite{mohap} and \cite{king-albright}.

Although neutrino oscillations are not sensitive to the overall neutrino
mass scale, there are other processes that do depend on absolute masses.
For example, studies of the tritium beta decay spectrum put an upper
limit on the effective mass of the electron neutrino
\begin{equation}
m_{\nu_e} \equiv \sum_j |U_{ej}|^2 m_j \,,
\end{equation}
of about 2.5~eV~\cite{tritium}; in a three-neutrino model this implies
an upper limit of 2.5~eV on the heaviest neutrino~\cite{bwwnumass}.

The current limit on the magnitude of the $\nu_e$--$\nu_e$ element of
the neutrino mass matrix for Majorana neutrinos from neutrinoless double
beta ($0\nu\beta\beta$) decay~\cite{doi} is of order 0.5~eV~\cite{baudis},
taking into account the imprecise
knowledge of the nuclear matrix element and the sensitivity of a background fluctuation analysis\cite{private}.
For the three-neutrino model this implies
\begin{eqnarray}
|M_{\nu_e\nu_e}| &=& |\sum_j U_{ej} m_j U_{ej}|
\\
&=& |(c_{13}c_{12})^2 m_1 + (c_{13}s_{12})^2 m_2 e^{i\phi_2}
+s_{13}^2 m_3 e^{i\phi_3}| \le M_{max} = 0.2{\rm~eV} \,,\nonumber 
\label{eq:Mee}
\end{eqnarray}
where $\phi_2$ and $\phi_3$ are extra phases present for Majorana
neutrinos that are not observable in neutrino oscillations. For models
with one large mass $m_1, m_2 \ll m_3 \simeq \sqrt{\delta m^2_{atm}}
\simeq 0.05$~eV, and Eq.~(\ref{eq:Mee}) does not provide any additional
constraint. However, if all three masses are nearly degenerate
($m_1 \simeq m_2 \simeq m_3 \equiv m$), the $0\nu\beta\beta$ decay limit
becomes $s_{12}^2 \ge [1 - 2s_{13}^2 - (M_{max}/m)]/(2c_{13}^2)$, which
in turn implies that the solar $\nu_e\to\nu_e$ oscillation amplitude
(see Eq.~(\ref{eq:P3solar})) has the constraint~\cite{bwmaj}
\begin{equation}
4c_{13}^4s_{12}^2c_{12}^2 \ge 1 - \left( {M_{max}\over m} \right)^2
-2s_{13}^2 \left( 1 + {M_{max}\over m} \right) \,.
\end{equation}
For any value of $m > M_{max}/(1-2s_{13}^2)$ there will be a lower limit
on the size of the solar $\nu_e\to\nu_e$ oscillation amplitude; for
example, given the current limit on $s_{13}$, the small-angle MSW solar
solution is ruled out for nearly degenerate Majorana neutrinos if
$m>0.25$~eV~\cite{bwmaj}. Large-angle MSW and vacuum solar solutions,
which have large mixing, are still allowed; any solar solution with
maximal mixing can never be excluded by this bound.

Neutrino mass also provides an ideal hot dark matter
component~\cite{hdm}; the contribution of neutrinos to the mass density
of the universe is given by $\Omega_\nu = \sum m_\nu/(h^2 93$~eV), where
$h$ is the Hubble expansion parameter in units of
100~km/s/Mpc\cite{expansion}. For $h=0.65$, the model with three nearly
degenerate neutrinos has $\Omega_\nu \simeq 0.075 (m/$eV). In
three-neutrino models with hierarchical masses, in which the largest
mass is of order $\sqrt{\delta m^2_{atm}}$, $\Omega_\nu$ is much
smaller, on the order of 0.001. Future measurements of the hot dark
matter component may be sensitive to $\sum m_\nu$ down to
0.4~eV~\cite{eht}.

\subsection{Four-neutrino models}

As described in Sec.~3.2, four-neutrino models must have the $2+2$
mass structure, i.e., two nearly-degenerate pairs separated from each
other by the LSND scale. One possible class of mass matrices that can
give this situation is
\begin{equation}
M = m \left( \begin{array}{cccc}
\epsilon_1 & \epsilon_2 & 0          & 0\\
\epsilon_2 & 0          & 0          & \epsilon_3\\
0          & 0          & \epsilon_4 & 1\\
0          & \epsilon_3 & 1          & \epsilon_5
\end{array} \right) \,,
\label{eq:m4}
\end{equation}
presented in the ($\nu_s,\nu_e,\nu_\mu,\nu_\tau$) basis (i.e., the basis
in which the charged lepton mass matrix is diagonal).

When $\epsilon_5=\epsilon_4$, the mass matrix in Eq.~(\ref{eq:m4}) can
accommodate any of the three solar solutions, depending on the hierarchy
of the mass matrix elements~\cite{4nuvar}
\begin{eqnarray}
{\rm SAM}: & \epsilon_2 \ll \epsilon_1, \epsilon_4 \ll \epsilon_3
\ll 1 \, \,,
\label{eq:ordera}\\
{\rm LAM}: & \epsilon_1, \epsilon_2, \epsilon_4 \ll \epsilon_3 \ll
1 \, \,,
\label{eq:orderb}\\
{\rm VO}: & \epsilon_1 \ll \epsilon_2 \ll \epsilon_4 \ll \epsilon_3 \ll
1 \,.
\label{eq:orderc}
\end{eqnarray}
In each case, the mass eigenvalues have the hierarchy $m_1 < m_0 \ll
m_2, m_3$, as required for the MSW solution. The mass matrix contains
five parameters, just enough to incorporate the required three
mass-squared differences and the oscillation amplitudes for the solar
and LSND neutrinos. The large amplitude for atmospheric oscillations
does not require an additional parameter, since the mass matrix
naturally gives nearly-maximal mixing of $\nu_\mu$ with $\nu_\tau$. For
the VO case, $\epsilon_1$ does not contribute to the phenomenology
and can be set to zero, so that there are only four parameters; the
large amplitude for solar oscillations also occurs naturally in this
case~\cite{4nuvar,roy}.

Another variant with five parameters is $\epsilon_5 = -\epsilon_4$ and
$\epsilon_2 \ll \epsilon_1 \ll \epsilon_3 \ll \epsilon_4 \sim
1$~\cite{gibbons}. In this case, $\epsilon_3$ determines both the
amplitude of the LSND oscillations {\it and} causes the splitting
between $m_2$ and $m_3$ that drives the atmospheric oscillations. Two
testable consequences of this model are that $\delta m^2_{atm} \le
1.3\times10^{-3}$~eV$^2$ and that there should be observable
$\nu_e\to\nu_\tau$ oscillations in short-baseline experiments.

For both of the previous cases ($\epsilon_5 = \epsilon_4 \ll 1$ and
$-\epsilon_5 = \epsilon_4 \sim 1$), the heavier mass pair is much
heavier than the lighter mass pair, i.e., $m_1 < m_0 \ll m_2, m_3$.  Yet
another possibility is to have $\epsilon_1 = \epsilon_5 = 0$ and
$\epsilon_3, \epsilon_4 \ll \epsilon_2 < 1$, where $\epsilon_2$ is not
small compared to unity~\cite{roy}. In this case, the two degenerate
pairs of masses have mass eigenvalues that are the same order of
magnitude; there are only four parameters as both the solar and
atmospheric mixings are naturally close to maximal. However, the
$m_0$--$m_1$ splitting in this model can give only MSW solar
oscillations, which for large mixing are disfavored when
$\nu_e\to\nu_s$.

Other four-neutrino mass matrix ansatzes have been presented in the
literature~\cite{other4nu}, but they generally have characteristics similar
to those discussed here.  In all of the explicit four-neutrino models
discussed above, since the $\nu_e$--$\nu_e$ element of the neutrino mass
matrix vanishes, there is no contribution to neutrinoless double beta
decay. However, because $m_3$ and $m_4$ are always of order 1~eV or more
(to provide the necessary mass-squared splitting for LSND
oscillations), these models always contribute to the hot dark matter in
the universe\cite{primack}. 

\section{Long-baseline experiments}

Long-baseline experiments (with $L/E_\nu \sim 10^2$--$10^3$~km/GeV) are
expected to confirm the $\nu_\mu\to\nu_\mu$ disappearance oscillations
at the $\delta m^2_{\rm atm}$ scale. The K2K experiment\cite{k2k} from
KEK to SuperK, with a baseline of $L\simeq 250$~km and a mean neutrino
energy of $\left< E_\nu\right> \sim 1.4$~GeV is underway. The MINOS
experiment from Fermilab to Soudan\cite{minos}, with a longer baseline
$L\simeq 730$~km and higher mean energies $\left< E_\nu \right> = 3$, 6
or 12~GeV, is under construction and the ICANOE\cite{icanoe} and
OPERA\cite{opera} experiments, with similar baselines from CERN to Gran
Sasso, have been approved.  These experiments with dominant $\nu_\mu $
and $\bar\nu_\mu$ beams will securely establish the oscillation
phenomena and may measure $\delta m^2_{\rm atm}$ to a precision of order
10\%\cite{thesis}. They will also measure the dominant oscillation
amplitude $\sin^22\theta_{23}$.

Further tests of the neutrino parameters will likely require higher
intensity neutrino beams, and $\nu_e$, $\bar\nu_e$ beams as well as
$\nu_\mu$, $\bar\nu_\mu$, such as those generated in a neutrino
factory~\cite{deruj,nufact1,nufact2,nufact3}. The $\nu_e$, $\bar\nu_e$
components of the beam allow one to search for $\nu_e\to\nu_\mu$ and
$\bar\nu_e\to\bar\nu_\mu$ appearance, which will occur in the leading
$\delta m^2_{atm}$ oscillation if $\sin^22\theta_{13}\ne0$.  Depending
on the machine parameters, $\delta m^2_{\rm atm}$ and
$\sin^22\theta_{23}$ can be measured to an accuracy of 1--2\%, and
$\sin^22\theta_{13}$ can be tested down to 0.01 or below~\cite{nufact3}.
If the baseline is long enough ($L \agt 1000$~km), matter effects in
the Earth will also play an important role in an appearance experiment:
for $\delta m^2_{\rm atm} > 0$ ($\delta m^2_{\rm atm} < 0$),
$\nu_e\to\nu_\mu$ oscillations are enhanced (suppressed) and
$\bar\nu_e\to\bar\nu_\mu$ oscillations are suppressed
(enhanced). Therefore by comparing $\nu_e\to\nu_\mu$ with
$\bar\nu_e\to\bar\nu_\mu$ oscillations it may be possible to determine
the sign of $\delta m^2_{\rm atm}$~\cite{nufact3}. The enhancement due
to matter of either $\nu_e\to\nu_\mu$ or $\bar\nu_e\to\bar\nu_\mu$ may
also improve the $\sin^22\theta_{13}$ sensitivity for appropriate
choices of $L$ and $E_\nu$~\cite{nufact3}. In a four-neutrino model,
both short-baseline experiments with $L/E_\nu \sim 1$~km/GeV (which
probe $\delta m^2_{\rm LSND}$) and long-baseline experiments will be
required to obtain maximal information on the neutrino mixing
parameters~\cite{4nuCP}.

$CP$-violating effects due to the phase $\delta$ only become
appreciable at the subleading $\delta m^2$ scale, and only then if the
mixing angles are large enough~\cite{CP3}. In a three-neutrino model
with $\delta m^2_{\rm sun}$ and $\delta m^2_{\rm atm}$, $CP$ violation
will be observable only for the large-angle MSW solar solution; a
long-baseline experiment with a high-intensity neutrino beam from a
neutrino factory may be able to give information on $\delta$ in this
case~\cite{CPlong}. In a four-neutrino model, potentially large
$CP$-violating effects are possible at the $\delta m^2_{\rm atm}$
scale~\cite{4nuCP,CPlarge}.

\section{Summary and outlook}

In a three-neutrino world, it may be possible to completely determine
the neutrino mixing matrix and two independent mass-squared differences
using existing and planned neutrino oscillation experiments.  Future
measurements of solar neutrinos should pin down the neutrino mass and
mixing parameters $\delta m^2_{21}$ and $\sin^22\theta_{12}$ that are
predominantly responsible for the solar neutrino deficit.
Long-baseline experiments can provide a more precise determination of
$\delta m^2_{32}$ and $\sin^22\theta_{23}$, which drive the atmospheric
neutrino anomaly, measure the size of $\sin^22\theta_{13}$, and determine the sign of $\delta m_{32}^2$. If the
solar solution is large-angle MSW, long-baseline experiments may also
be sensitive to the $CP$-violating phase $\delta$. Future measurements
of beta decay spectra, double-beta decay and hot dark matter may then
help determine the last remaining neutrino mass parameter, the absolute
neutrino mass scale.

In a four-neutrino world, there are three additional mixing angles and
three additional $CP$ phases. Since the extra neutrino is sterile, it
may be difficult to determine some of the additional mixing angles,
especially if they are small, as is the case in many existing models.
Short baseline experiments that probe the $\delta m^2_{\rm LSND}$
oscillation will be helpful in making sense of the larger parameter
space. Furthermore, in four-neutrino models $CP$ violation may become
observable at the $\delta m^2_{\rm atm}$ scale (rather than the $\delta
m^2_{\rm sun}$ scale, as is the case in three-neutrino models), and
there will be a contribution to hot dark matter on the order of $\sum
m_\nu \sim 1$~eV.

\end{document}